\documentstyle[12pt]{article}

\input epsf

\newcommand{\resection}[1]{\setcounter{equation}{0}\section{#1}}

\newcommand{\beqn}{\begin{equation}}
\newcommand{\eeqn}{\end{equation}}
\newcommand{\beqa}{\begin{eqnarray}}
\newcommand{\eeqa}{\end{eqnarray}}
\newcommand{\s}{\sigma}
\newcommand{\la}{\lambda}
\renewcommand{\thefootnote}{\fnsymbol{footnote}}

\begin{document}
\begin{flushright}

        {\normalsize      OU-HET 224\\
                         October, 1995\\}

\end{flushright}

\vfill
\begin{center}
{\large {\bf Functional Equations of Form Factors \\
for Diagonal Scattering Theories}}
\vfill

{\bf Takeshi Oota}\footnote{This work was supported by
JSPS Research Fellow.}

        Department of Physics,  Faculty of Science, \\
        Osaka University,  Toyonaka, Osaka 560, Japan \\

\end{center}
\vfill
\renewcommand{\thefootnote}{\arabic{footnote}}
\setcounter{footnote}{0}

\begin{abstract}

Form factor bootstrap approach is applied for diagonal scattering theories.
We consider the ADE theories and determine the functional equations
satisfied by the minimal two-particle form factors. We also determine the
parameterization of the singularities in two particle form factors.

For $A^{(1)}_{2}$ Affine Toda field theory which is the simplest non-self
conjugate theory, form factors are derived up to four-body and
identification of operator is done.
Generalizing this identification to the $A^{(1)}_N$ Affine Toda cases,
we fix the two particle form factors. We also determine the additional
pole structure of form factors which comes from the double
pole of the $S$-matrices of the $A^{(1)}_N$ theory.

For $A_N$ theories, existence of the conserved ${\bf Z}_{N+1}$ charge leads to
the division of the set of form factors into $N+1$ decoupled sectors.

\end{abstract}
\vfill

\section{Introduction}

For two dimensional factorizable scattering theories,
the bootstrap framework gives strong constraints on
physical quantities. For example, under some assumptions such as unitarity,
the bootstrap determines the $S$-matrices
almost completely and non-perturbatively.

The factorizable and diagonal scattering theories are integrable models
which have deep relationship with underlying Lie algebras.
The $S$-matrices for these theories are well known and determined from
the data of the associated Lie algebras.

Less-known objects for these theories are
form factors i.e. matrix elements of operators.
If we know all form factors then the correlation function can be
calculated in principle.

In integrable theories, the form factors can also be determined by
the form factor bootstrap approach \cite{KW,S}.

The form factors for the diagonal scattering theories are known for
the thermal perturbation of the Ising model \cite{CM,YZ},
the scaling Lee-Yang model \cite{Z},
the sinh-Gordon model \cite{KM,FMS2},
the Bullough-Dodd model ($A^{(2)}_2$) \cite{FMS} and
the $\phi_{1,3}$-perturbed non-unitary $M_{3,5}$ model \cite{DM2}
all of which contain only one type of particle,
and for minimal $A^{(2)}_{2N}$ theories which contain $N$ kinds of particles
\cite{S2}.
Two-body form factors for the magnetic perturbation of
the Ising model can be found in \cite{DM}.

Koubek shows that for minimal $A^{(2)}_{2N}$ theories,
($\phi_{1,3}$-perturbed $M_{2,2N+3}$ minimal conformal field theories),
all recursion relations for form factors can be simplified to the
recursion relations for the form factors which contain only one kind of
particles \cite{KN1}.

All particles (or excitations) in these theories are self-conjugate, i.e.
a particle and its anti-particle are identical.

But for the case of non self-conjugate theories, the kinematical residue
equations can not be used directly to determine the form factor.
In this case, recursion relations for one particle become rather
difficult to solve.
In this paper, we take a straightforward approach:
trying to solve simultaneously the system of recursion relations.

We mainly deal with the $A^{(1)}_N$ models, especially
the $A^{(1)}_2$ model.

The paper is organized as follows. In section 2.1 we briefly
review the form factor bootstrap approach to fix the notation.
In section 2.2, we discuss the properties of minimal
two particle form factors and determine the parameterization of
two particle form factors.
In section 3 we derive the form factors of  $A^{(1)}_2$ Affine Toda
field theory up to four-body. And we identify the fundamental operators.
The generalization to $A_N$ theories are discussed in section 4, and
we determine the two-particle form factors.
Section 5 are conclusions and discussion.

\resection{The form factor bootstrap}

\subsection{Equations for the form factors}

The matrix elements of a local operator ${\cal O}(x)$
\beqa
& &F^{a'_1a'_2\cdots a'_m}_{a_1a_2\cdots a_{n}}
(\beta'_1, \beta'_2, \cdots ,\beta'_m|\beta_1, \beta_2, \cdots ,\beta_n)
\nonumber \\
&=&^{a'_1a'_2\cdots a'_m}<\beta'_1, \beta'_2, \cdots ,\beta'_m|
{\cal O}(0)|\beta_1, \beta_2, \cdots ,\beta_n>_{a_1a_2\cdots a_n},
\label{gff}
\eeqa
are called form factors of general kind.
Here $\beta_i$ is a rapidity of a particle of a species $a_i$.
Consider the following form of matrix elements
\beqn
F_{a_1a_2\cdots a_{n}}(\beta_1, \beta_2, \cdots ,\beta_n)=
<0|{\cal O}(0)|\beta_1, \beta_2, \cdots ,\beta_n>_{a_1a_2\cdots a_n}.
\label{ff}
\eeqn
The general form factors (\ref{gff}) are related to the functions (\ref{ff})
by analytic continuation
\beqa
& &F^{a'_1\cdots a'_m}_{a_1\cdots a_{n}}
(\beta'_1, \cdots ,\beta'_m|\beta_1, \cdots ,\beta_n)
\\
&=&{\cal C}^{a'_1b_1}\cdots {\cal C}^{a'_mb_m}F_{b_1\cdots b_ma_1\cdots a_{n}}
(\beta'_1+i\pi, \cdots ,\beta'_m+i\pi, \beta_1, \cdots ,\beta_n), \nonumber
\eeqa
provided that the set of rapidities $\beta'$ are separated from the set
$\beta$. Here ${\cal C}^{ab}$ is inverse of charge conjugation matrix
${\cal C}_{ab}$ and for ADE scattering theories
${\cal C}_{ab}=\delta_{a\bar{b}}$.

Watson's equations for diagonal scattering theories take a simple form
\beqa
& &F_{a_1\cdots a_ia_{i+1}\cdots a_n}
(\beta_1, \cdots, \beta_i, \beta_{i+1}, \cdots, \beta_n) \nonumber \\
&=&S_{a_ia_{i+1}}(\beta_i-\beta_{i+1})F_{a_1\cdots a_{i+1}a_{i}\cdots a_n}
(\beta_1, \cdots, \beta_{i+1}, \beta_i, \cdots, \beta_n), \nonumber
\eeqa
\beqn
F_{a_1a_2\cdots  a_n}(\beta_1+2\pi i, \beta_2, \cdots, \beta_n)=
F_{a_2\cdots  a_na_1}(\beta_2, \cdots, \beta_n, \beta_1).
\eeqn

The simple pole structure of the form factors are summarized by the following
two types of the recursion relations.\footnote{
If $S$-matrices contain double poles then additional simple pole
structure appears.}

The first kind of relations are called kinematical residue equation
\beqn
-i \ {\rm res}_{\beta'=\beta+i\pi} \ F_{\bar{a}ad_1\cdots d_n}
(\beta', \beta ,\beta_1, \cdots, \beta_n)
=F_{d_1\cdots d_n}(\beta_1, \cdots, \beta_n)
\left(1-\prod_{j=1}^nS_{ad_j}(\beta-\beta_j)\right).
\eeqn
And the second kind of relations are called
bound state residue equation
\beqn
-i \ {\rm res}_{\beta'=\beta+i\theta_{ab}^c} \ F_{abd_{1}\cdots d_{n}}
(\beta', \beta,\beta_1, \cdots, \beta_n)
=
\Gamma_{ab}^cF_{\bar{c}d_1\cdots d_n}
(\beta+i\bar{\theta}_{bc}^a, \beta_1, \cdots, \beta_n),
\eeqn
where $\bar{\theta}=\pi-\theta$.
The on-shell three point vertex $\Gamma_{ab}^c$ is given by
\[
-i \ {\rm res}_{\beta=i\theta_{ab}^c} \ S_{ab}(\beta)=
(\Gamma_{ab}^{c})^{2}.
\]

\subsection{Minimal form factors}

In the case of $n=2$, Watson's equations reduce to
\[
F_{ab}(\beta)=S_{ab}(\beta)F_{ba}(-\beta),
\]
\beqn
F_{ab}(\beta+2\pi i)=F_{ba}(-\beta).
\label{Wat}
\eeqn

The general solution of Watson's equations takes the form \cite{KW}
\beqn
F_{a_1\cdots a_n}(\beta_1, \cdots, \beta_n)=
K_{a_1\cdots a_n}(\beta_1, \cdots, \beta_n)
\prod_{i<j}F^{(min)}_{a_ia_j}(\beta_i-\beta_j),
\label{FKf}
\eeqn
where $F^{(min)}_{ab}$ is solution of eq.(\ref{Wat}) which is analytic in
the strip $0\leq Im\beta\leq 2\pi$
and has no zeros in $0<Im\beta<2\pi$.

The building block of diagonal S-matrices is
$(x)_{\beta}=(x)_{+(\beta)}/(-x)_{+(\beta)}$ where
$(x)_{+(\beta)}=\frac{1}{i\pi}\sinh\frac{1}{2}(\beta+i\frac{\pi}{h}x)$.
We write the basic building block of minimal (two particle)
form factor corresponding to
$(x)_{\beta}$ as following forms
\beqn
f_{x}(\beta)=\sinh\frac{1}{2}\beta\frac{g_x(\beta)}{g_{2h-x}(\beta)}
\eeqn
which has no poles and zeros in the strip $0<Im\beta<2\pi$ for $0\leq x<2h$.
The function $g_x(\beta)$ is given by
\beqn
g_{x}(\beta)\equiv\prod_{n=1}^{\infty}
\frac{\Gamma(n+\frac{i\beta}{2\pi}-\frac{x}{2h}+1)}
{\Gamma(n-\frac{i\beta}{2\pi}+\frac{x}{2h}-1)}.
\eeqn
The function $g_{x}(\beta)$ has poles at $\beta=-i\pi x/h+2(m+2)\pi i$ for
$m=0,1,2,\ldots$ and zeros at $\beta=-i\pi x/h-2m\pi i$ for
$m=0,1,2,\ldots$.

The introduction of the function $g_x$ simplifies the calculation.
Using the following properties of $g_{x}$
\beqn
g_{x}\left(\beta+i\frac{\pi}{h}y\right)=g_{x+y}(\beta),
\eeqn
\beqn
g_{x}(\beta+2\pi i)=g_{x+2h}(\beta)=\frac{1}{(x)_+}g_{x}(\beta),
\label{dim}
\eeqn

\beqn
g_{x}(i\pi -\beta)=\frac{1}{g_{3h-x}(\beta)}=
\frac{(h-x)_+}{g_{h-x}(\beta)},
\eeqn

\beqn
g_x(0)g_{4h-x}(0)=1,
\eeqn
the behavior of the minimal form factors under the recursion
relation is easily determined.

For ADE scattering theories,
the basic building blocks of the diagonal $S$-matrices are
$<x>_{\beta}=<x>_{+(\beta)}/<-x>_{+(\beta)}$
with
\[
<x>_+=\left\{\begin{array}{lll}
(x-1)_+(x+1)_+ & {\rm for} \ {\rm perturbed \ conformal} \\
 & \\
{(x-1)_+(x+1)_+ \over {\strut (x-1+B)_+(x+1-B)_+}} &
{\rm for} \ {\rm Affine \ Toda}\end{array}\right.
\]
where $h$ is the Coxeter number of the associated Lie algebra.

So the building block corresponding to $<x>$ is
\beqn
F_x^{(min)}(\beta)=\frac{G_x(\beta)}{G_{2h-x}(\beta)},
\eeqn
where
\[
G_{x}(\beta)=\left\{
\begin{array}{lll}
g_{x-1}(\beta)g_{x+1}(\beta) & {\rm for} \ {\rm perturbed \ conformal} \\
 & \\
{g_{x-1}(\beta)g_{x+1}(\beta)\over{\strut g_{x-1+B}(\beta)g_{x+1-B}(\beta)}} &
{\rm for} \ {\rm Affine \ Toda}.\end{array}\right.
\]

We list some properties of $G_{x}$.
\beqn
G_{x}\left(\beta+i\frac{\pi}{h}y\right)=G_{x+y}(\beta),
\eeqn
\beqn
G_{x}(\beta+2\pi i)=G_{x+2h}(\beta)=\frac{1}{<x>_{+}}G_{x}(\beta),
\label{dim2}
\eeqn
\beqn
G_{x}(i\pi -\beta)=\frac{1}{G_{3h-x}(\beta)}=
\frac{<h-x>_{+}}{G_{h-x}(\beta)}.
\eeqn
\beqn
G_x(0)G_{4h-x}(0)=1.
\eeqn
If $S$-matrices have the following form
\beqn
S_{ab}(\beta)=\prod_{x\in A_{ab}}<x>_{\beta},
\eeqn
then the minimal solutions of eq.(\ref{Wat}) are written as
\beqn
F^{(min)}_{ab}(\beta)=\prod_{x\in A_{ab}}F_{x}^{(min)}(\beta).
\label{mFF}
\eeqn

Here $A_{ab}$ is the minimal set of numbers which gives the correct
$S_{ab}$. The multiplicity of $p$ in $A_{ab}$ is denoted by
$m_p(A_{ab})$. The sets $A_{ab}$ are chosen such that
$m_{p}(A_{ab})=0$ for $p\leq 0$ or for $p\geq h$.

The crossing condition of the $S$-matrix
$S_{ab}(i\pi-\beta)=S_{b\bar{a}}(\beta)$ is
equivalent to the condition $m_{h-p}(A_{ab})=m_p(A_{b\bar{a}})$.

The minimal form factor (\ref{mFF}) has following
property
\beqn
F^{(min)}_{\bar{a}b}(\beta+i\pi)F^{(min)}_{ab}(\beta)=1/\xi_{ab}(\beta).
\label{ffu}
\eeqn
Eq.(\ref{ffu}) is an analogue of the $S$-matrix relation
$S_{\bar{a}b}(\beta+i\pi)S_{ab}(\beta)=1$.
There is one-to-one correspondence between the constituent of
minimal form factors and the $S$-matrices.
Many properties of $F_x^{(min)}$ are similar to those of $<x>$
but the monodromy property is quite different.
$F_x^{(min)}$ has a diagonal monodromy which comes from eq.(\ref{dim2})
while $<x>$ is $2\pi i$-periodic.
Due to these monodromy factors,
additional functions $\xi_{ab}(\beta)$ appear in the
right hand side of eq.(\ref{ffu}).
The factor $1/\xi_{ab}$ is given by the product of $<x>_+$
\beqn
1/\xi_{ab}(\beta)=\prod_{x\in A_{ab}}<x>_{+(\beta)}.
\label{Bab}
\eeqn
In particular, for Affine cases,
$<x>_+=(x-1)_+(x+1)_+/(x-1+B)_+(x+1-B)_+$ implies that $B$ dependent parts
appear only in the numerator of $\xi_{ab}$.
The form of $\xi$ has been determined for the scaling Lee-Yang model \cite{Z},
the sinh-Gordon model \cite{FMS2}
and for the Bullough-Dodd model \cite{FMS} by explicit calculation.

Using eq. (\ref{Wat}), we get the following relations:
\beqn
S_{ab}(\beta)=\frac{\xi_{\bar{a}b}(\beta+i\pi)}{\xi_{ab}(\beta)}=
\frac{\xi_{ab}(-\beta)}{\xi_{ab}(\beta)} .
\eeqn
Under the factorization of eq.(\ref{FKf}),
the kinematical residue equations reduce to
\beqa
\label{KKR}
& &-i \ {\rm res}_{\beta'=\beta+i\pi} \
K_{\bar{a}ad_{1}\cdots d_{n}}
(\beta', \beta ,\beta_{1}, \cdots, \beta_{n})\\
&=&K_{d_{1}\cdots d_{n}}(\beta_{1}, \cdots, \beta_{n})
\left(\prod_{j=1}^{n}\xi_{ad_{j}}(\beta-\beta_{j})-
\prod_{j=1}^{n}\xi_{ad_{j}}(\beta_{j}-\beta)\right)/F^{(min)}_{\bar{a}a}(i\pi).
\nonumber
\eeqa

In accordance with the $S$-matrix bootstrap
\beqn
S_{ad}(\beta+i\bar{\theta}^b_{ac})S_{bd}(\beta-i\bar{\theta}^a_{bc})=
S_{\bar{c}d}(\beta),
\eeqn
the minimal form factors have following properties
\beqn
F^{(min)}_{ad}(\beta+i\bar{\theta}^b_{ac})
F^{(min)}_{bd}(\beta-i\bar{\theta}^a_{bc})=
F^{(min)}_{\bar{c}d}(\beta)/\la^c_{ab;d}(\beta).
\eeqn
The extra factor $1/\la^c_{ab;d}(\beta)$ comes from the diagonal monodromy
and is given by the product of $<x>_+$:
\beqn
1/\la^c_{ab;d}(\beta)=\prod_{\{x\in A_{ad}|x<\bar{u}_{ac}^b\}}
<\bar{u}_{ac}^b-x>_{+(\beta)}
\prod_{\{x\in A_{bd}|x<\bar{u}_{bc}^a\}}<x-\bar{u}_{bc}^a>_{+(\beta)}.
\eeqn
Here $u^c_{ab}=h\theta^c_{ab}/\pi$ and $\bar{u}^c_{ab}=h-u^c_{ab}$.

We list some properties of $\la_{ab;d}^c$:
\[
\la_{ab;d}^c(\beta)=\la_{ba;d}^c(-\beta),
\]
\[
\la_{ca;\bar{d}}^b(\beta-i\theta_{ab}^c)
\la_{ab;d}^{c}(\beta-i\bar{\theta}_{ac}^b)=\xi_{ad}(-\beta).
\]

The function $\la$ is known for the scaling Lee-Yang model \cite{Z} and
the Bullough-Dodd model \cite{FMS,FFFA}.

The Bound state residue equations reduce to
\beqa
& &-i \ {\rm res}_{\beta'=\beta+i\theta_{ab}^c} \
K_{abd_{1}\cdots d_{n}}
(\beta', \beta,
\beta_1, \cdots, \beta_n) \nonumber \\
&=&
\Gamma_{ab}^cK_{\bar{c}d_1\cdots d_n}
(\beta+i\bar{\theta}_{bc}^a, \beta_1, \cdots, \beta_n)
\prod_{j=1}^n\la_{ab;d_j}^c(\beta+i\bar{\theta}_{bc}^a-\beta_j)
/F^{(min)}_{ab}(i\theta_{ab}^c).
\label{KBR}
\eeqa

For ADE scattering theories, the $S$-matrix can be written as \cite{D}
\beqn
S_{ab}(\beta)=
\prod_{p=0}^{h-1}\left(<2p+1+\epsilon_{ab}>_{+(\beta)}\right)
^{\mu^{(a)}\cdot w^{-p}\phi_b}.
\eeqn
Or equivalently
\[
m_x(A_{ab})=\mu^{(a)}\cdot w^{-p}\phi_b \ \ \
{\rm for}\ \  0<x=2p+1+\epsilon_{ab}<h.
\]
Here $\mu^{(a)}$ is the fundamental weight of the algebra.
So the minimal form factors can be given by\footnote{
For the minimal cases, if the $S$-matrix have fermionic nature:
$S_{ab}(0)=-1$, then one more factor $(0)_{+(\beta)}$ is needed.}
\beqn
F^{(min)}_{ab}(\beta)=<0>_+^{\delta(\epsilon_{ab},1)\mu^{(a)}\cdot w\phi_b}
\prod_{p=0}^{h-1}(G_{2p+1+\epsilon_{ab}}(\beta))
^{\mu^{(a)}\cdot w^{-p}\phi_b}.
\label{mffADE}
\eeqn
Here $\epsilon_{ab}=\frac{1}{2}(c(a)-c(b))$.
Depending the two colourings of the Dynkin diagram of the
algebra associated,
$c(a)=1$ for white nodes and $c(a)=-1$ for black nodes \cite{FLO}.

Note that we can see $c(\bar{a})=(-1)^hc(a)$.
If the Coxeter number $h$ is even, which holds except for $A_{2N}$ theories,
$\epsilon_{\bar{a}b}=\epsilon_{ab}$.

The first adjustment factor in the right hand side of eq. (\ref{mffADE})
is introduced in order that $F^{(min)}_{ab}$ is
constructed from $G_x \ (0\leq x<2h)$ for $\epsilon_{ab}=1$.

In the right hand side of eq.(\ref{KKR}) and eq.(\ref{KBR}),
for the Affine cases,
coupling dependent parts $(x\pm B)_+$ only appear in positive powers,
so the singularities of $K$ can be factorized by products of $1/(x)_+$.

\beqn
K_{a_1\cdots a_n}(\beta_1,\cdots,\beta_n)=
Q_{a_1\cdots a_n}(\beta_1,\cdots,\beta_n)\prod_{i<j}
\frac{1}{(e^{\beta_i}+e^{\beta_j})^{{\cal C}_{a_ia_j}}
W_{a_ia_j}(\beta_i-\beta_j)}.
\eeqn
The function $W_{ab}$ contains the factor $(u^c_{ab})_+(-u^c_{ab})_+$
for bound state poles.
The factor $(-u^c_{ab})_+$ is needed to make $W_{ab}$ symmetric:
$W_{ba}(-\beta)=W_{ab}(\beta)$.
In general, $W_{ab}$ must contain more factors to factorize
the higher order poles in $\xi_{ab}$.

The polynomial $Q_{a_1\cdots a_n}$ carries the information about
operators. Counting the number of independent solution
$Q_{a_1\cdots a_n}$ can be used to classify the operator content of the
model \cite{CM,KM,KN1,KN2,ScS}.

We expect that $Q_{a_1,\cdots,a_n}$ are polynomials in $(x)_+(-x)_+$.
So the function $W_{ab}$ is determined from the requirement that
$W_{\bar{a}b}(\beta+i\pi)W_{ab}(\beta)\xi_{ab}(\beta)$ is product of
the factors $(x)_+$ in positive powers.

Using the expression (\ref{Bab}), we write the singularity of $\xi_{ab}$ as
follows
\[
\prod_{x\in A_{ab}}\frac{1}{(x-1)_+(x+1)_+}
=\prod_{p=0}^{h}(p)_+^{-m_{p-1}(A_{ab})-m_{p+1}(A_{ab})}.
\]

Even order poles do not correspond to the bound state.
If $\xi_{ab}$ contain the factor $1/(x)_+^{2k}$,
the half $1/(x)_+^k$ is canceled by
$W_{ab}(\beta)$ and the other half is canceled by $W_{\bar{a}b}(\beta+i\pi)$.
Then $W_{ab}(\beta)$ must contain $(x)_+^k(-x)_+^k$,
and $W_{\bar{a}b}(\beta)$ have $(x-h)_+^k(h-x)_+^k$.

Odd order poles can be interpreted as the production of a bound
state. $W_{ab}$ contains the factor $(u_{ab}^c)_+(-u_{ab}^c)_+$
for the bound state in forward channel.
If $\xi_{ab}$ contains
the factor $1/(x)_+^{2k+1}$ then $W_{ab}$ has the factor $(x)_+^{k+1}$
and $W_{\bar{a}b}$ has the factor $(x-h)_+^{k}$ for the forward channel.

Dorey's ``uphill/downhill'' mnemonic \cite{D} implies that
$m_{p-1}-m_{p+1}=+1,0,-1$ and the case $m_{p-1}-m_{p+1}=+1$ corresponds to
the forward channel.

The above consideration leads to the following form of the parameterization:
\[
W_{ab}(\beta)=
\rho_{ab}\left(\beta+i\frac{\pi}{h}\right)
\rho_{ab}\left(-\beta+i\frac{\pi}{h}\right),
\]
where
\[
\rho_{ab}(\beta)=\prod_{\{x\in A_{ab}|x\neq h-1\}}(x)_{+(\beta)}.
\]

The singularity structure of $\xi_{ab}$ is similar to that of $S_{ab}$.
The singularities of the $S$-matrices for ADE theories are explained
in terms of multi-scattering processes \cite{CT,BCDS}. So it is natural to
expect that the factorization in the above admits such interpretation.

For perturbed conformal theories,
Delfino and Mussardo \cite{DM} have derived the parameterization of
the singularities of two particle form factors from
the Feynman diagrammatic analysis of multi-particle processes.
Their factorization is agrees with our result.

\resection{$A^{(1)}_2$ case}

The $A^{(1)}_{2}$ theory is the simplest model based on the Lie algebra which
contain non self-conjugate particles.
It contains two kinds of particles, which are denoted by $1$ and $2$.
The particle $2$ is the  anti-particle of $1$ and vice versa.

The $S$-matrices of this theory are given by $S_{11}=S_{22}=<1>$
and $S_{12}=S_{21}=<2>$ \cite{BCDS,AFZ,KlM}.
For definiteness, we consider the Affine $A^{(1)}_{2}$ theories.
We define
\beqa
& &F_{[m,n]}(\beta_{1}, \beta_{2}, \cdots,
\beta_{m}; \beta'_{1}, \beta'_{2}, \cdots, \beta'_{n}) \nonumber \\
&=&
F_{11\cdots 122\cdots 2}(\beta_{1}, \beta_{2}, \cdots,
\beta_{m}, \beta'_{1}, \beta'_{2}, \cdots, \beta'_{n}).
\eeqa
And we factorize $K_{[m,n]}$ as follows
\beqa
& &K_{[m,n]}(\beta_1,\cdots,\beta_m;\beta'_1,\cdots,\beta'_n) \\
&=&
{Q_{[m,n]}(x_1,\cdots,x_m;y_1,\cdots,y_n)\over
{\strut \prod_{i<j}(x_i-\omega^2 x_j)(x_i-\omega^{-2} x_j)
\prod_i\prod_j(x_i+y_j)
\prod_{i<j}(y_i-\omega^2 y_j)(y_i-\omega^{-2} y_j)}}
\nonumber
\eeqa
with $x_i=e^{\beta_i}$, $y_i=e^{\beta'_i}$ and $\omega=e^{i\frac{\pi}{3}}$.
The degree of polynomial $Q_{[m,n]}$ is given by
$deg(Q_{[m,n]})=(m+n)(m+n-1)-mn$.
For simplicity, we use the vector notation ${\bf x}=(x_1,x_2,\cdots)$ etc.

Then the kinematical residue equation is reduced to
\beqn
Q_{[m+1,n+1]}({\bf x},-x;x,{\bf y})=x D_{[m,n]}(x;{\bf x},{\bf y})
Q_{[m,n]}({\bf x};{\bf y}),
\label{kinQ}
\eeqn
where
\beqa
& &D_{[m,n]}(x;{\bf x},{\bf y})=(-1)^{n}H_2 \nonumber \\
& &\times\left(D_m(x;{\bf x};-\omega)D_n(x;{\bf y};\omega^{-1})
-D_m(x;{\bf x};-\omega^{-1})D_n(x;{\bf y};\omega)\right). \nonumber
\eeqa
The function $D_m$ is defined by
\beqa
\label{Dm}
D_m(x;{\bf x};z)&=&\prod_{j=1}^{m}(x+zx_j)(x-zqx_j)(x-zq^{-1}x_j)
\\
&=&\sum_{l=0}^m\sum_{k=0}^m\sum_{r=0}^{m-k}(-1)^k	[k+1]_q
x^{3m-2r-k-l}z^{2r+k+l}\s_l(\bf{x})s_{(2^r,1^k)}(\bf{x}), \nonumber
\eeqa
where $[k]_q=(q^k-q^{-k})/(q-q^{-1})$ and
the Schur function $s_{(2^r,1^k)}$ is
\[
s_{(2^r,1^k)}=(\s_{r+k}\s_{r-2}-\s_{r+k-1}\s_{r-1}).
\]
Here $\s_j$ are the elementary symmetric polynomials defined by
\[
\prod_{j=1}^m(x+x_j)=\sum_{j=0}^mx^{m-j}\s_j({\bf x}).
\]

The $B$ dependent parameter $H_2$ is defined by
\[
H_2=\frac{(-i)}{F^{(min)}_{12}(i\pi)}.
\]
The coupling dependent parameter $q=e^{i\frac{\pi}{3}(B-1)}$
transforms into $q^{-1}$ under the weak-strong
transformation $B\rightarrow 2-B$.

The bound state residue equations are reduced to
\[
Q_{[m+2,n]}({\bf x},\omega y,\omega^{-1}y;{\bf y})=H y^2
D_m(y;{\bf x};1)Q_{[m,n+1]}({\bf x};y,{\bf y}),
\]
\beqn
Q_{[m,n+2]}({\bf x};\omega x,\omega^{-1}x,{\bf y})=H x^2
D_m(x;{\bf y};1)Q_{[m+1,n]}({\bf x},x;{\bf y}),
\label{bouQ}
\eeqn
where
\[
H=-\frac{\sqrt{3}\Gamma}{F^{(min)}_{11}(\frac{2}{3}\pi i)}
=-\frac{\sqrt{3}\Gamma}{F^{(min)}_{22}(\frac{2}{3}\pi i)}.
\]
The function $\Gamma$ is given by
\beqn
(\Gamma)^2=(\Gamma_{11}^{1})^2=(\Gamma_{22}^{2})^2=
\sqrt{3}\frac{\sin\frac{\pi}{6}B\sin\frac{\pi}{6}(2-B)}
{\sin\frac{\pi}{6}(4-B)\sin\frac{\pi}{6}(2+B)}.
\eeqn
There is the relation between $H$ and $H_2$
\beqn
H^2/H_2=(1+\omega)(1+\omega [2]_q),
\eeqn
which is equivalent to the minimal form factor relation
\beqn
\left(F^{(min)}_{11}(\frac{2}{3}\pi i)\right)^2/F^{(min)}_{12}(i\pi)=
<3>_{+(0)}.
\eeqn

We first pay attention only to the indices $[m,n]$.
The bound state residue equations relate $Q_{[m,n]}$ to $Q_{[m-2,n+1]}$ or
$Q_{[m+1,n-2]}$ and the kinematical ones relate $Q_{[m,n]}$ to $Q_{[m-1,n-1]}$.
We identify $[m,n]$ with two dimensional vector and
introduce the following equivalence relations
\beqn
[m,n]\sim[m,n]-l_{1}[2,-1]-l_{2}[-1,2]-l_{3}[1,1]
\label{IEQ}
\eeqn
where $l_{i}\in {\bf Z}$.
Because $[1,1]$ are equal to $[2,-1]+[-1,2]$, the above
definition is redundant.
But for later convenience, we write here the $[1,1]$ term.

Then there are three equivalence classes
\beqn
\{[m,n]\}/\sim=[[0,0]]+[[1,0]]+[[0,1]].
\eeqn
In other words, if $[m,n]$ and $[m',n']$ belong to different classes then
$Q_{[m,n]}$ and $Q_{[m',n']}$ are not linked by residue equations.

So the set of form factors is divided into three sectors.
In each sector, higher polynomials are determined iteratively from
lower ones. In the minimal polynomial space, the solutions can be
determined uniquely except for the kernel ambiguity.

Note that $[2,-1]$ and $[-1,2]$ are equal to the first and second rows of
the Cartan matrix of $A_2$ algebra respectively.

The index $[m,n]$ can be identified with the Dynkin indices of weights
and with corresponding weights $\mu$
\beqn
\mu=[m_1, m_2]=m_1\mu^{(1)}+m_2\mu^{(2)},
\eeqn
where $\mu^{(a)}$ are the fundamental weights.
Then the equivalence relation can be rewritten as follows
\beqn
\mu\sim\mu-l_1\alpha_1-l_2\alpha_2-l_3(\alpha_1+\alpha_2).
\eeqn
Here $\alpha_1$ and $\alpha_2$ are the simple roots of the $A_2$ algebra.
For the $A^{(1)}_{2}$ theory, the equivalence relation is generated by
all positive roots.

In the polynomial space we are considering, the degree of a polynomial
is equal to the degree of the kernel and the kernel is one dimensional.
So at every recursion step, only one parameter enter to the solution space
and the parameter $A_{\nu}$ is attached to the point $\nu$ in the dominant
weight lattice.
The most general solution of the residue equations (\ref{kinQ}) and
(\ref{bouQ}) has following forms
\beqn
Q_{\mu}({\bf x},{\bf y})
=\sum_{\nu\leq\mu}H^{|\mu-\nu|}A_{\nu}Q_{\mu, \nu}({\bf x},{\bf y}).
\label{QHAQ}
\eeqn
Here the ordering of weights $\mu\geq\nu$ means that
the difference $\mu-\nu$ lies in the dominant root lattice and
$|\mu|=m_1+m_2$ for $\mu=[m_1,m_2]$.

The polynomial $Q_{\mu}$ carries the information about operators.
Each $Q_{\mu, \nu}$ satisfies the residue equations,
so gives an independent form factor of some operator ${\cal O}_\nu$.

\subsection{The $[0,0]$-sector}

We start from the $[0,0]$-sector to solve the recursion relations.

Although we call $[0,0]$-sector, the kinematical recursion relation is not
applied to the $Q_{[1,1]}\rightarrow Q_{[0,0]}$
\footnote{If the recursion process was started from $Q_{[0,0]}$ formally,
we would only get $Q_{[m,n]}=Q_{[0,0]}\delta_{m,0}\delta_{n,0}$.
This solution corresponds to the 'form factor of identity operator'.}.

So the first polynomial is  $Q_{[1,1]}$.
The most general degree $1$ polynomial is
\beqn
Q_{[1,1]}(x,y)=A_{[1,1]}x+A'_{[1,1]}y
\eeqn
where $A_{[1,1]}$ and $A'_{[1,1]}$ are constants.
But the recursion relation $Q_{[3,0]}\rightarrow Q_{[1,1]}$ has no solution
unless $A_{[1,1]}=A'_{[1,1]}$.
Similar phenomena occur at higher stages of recursion processes.
These additional constraints come from ${\bf Z}_2$ symmetry corresponding to
the charge conjugation. These ${\bf Z}_2$ constraints are imposed on
the constants: $A_{[m,n]}=A_{[n,m]}$.

First few solutions in the $[0,0]$-sector are given by
\begin{flushleft}
\[
Q_{[1,1]}(x,y)=A_{[1,1]}(x+y).
\]
\[
Q_{[3,0]}=A_{[3,0]}B_{1[3,0]}+HA_{[1,1]}\s_1\s_2
\left(\s_1\s_2-(2+[2]_q)\s_3\right).
\]
\[
Q_{[2,2]}=A_{[2,2]}B_{1[2,2]}K_{[2,2]}B_{2[2,2]}
+HA_{[3,0]}Q_{[2,2],[3,0]}+H^2A_{[1,1]}Q_{[2,2],[1,1]}.
\]
\end{flushleft}

Here
\beqa
Q_{[2,2],[0,3]}&=&
(\s_1^2(x)\s_1^2(y)-\s_2(x)\s_2(y)) \nonumber \\
&\times&((\s_2(x)-\s_2(y))^2+
\s_1(x)\s_1(y)(\s_2(x)+\s_2(y))+\s_2(x)\s_1^2(y)+\s_1^2(x)\s_2(y)), \nonumber
\eeqa

\beqa
Q_{[2,2],[1,1]}&=&
(\s_1(x)+\s_1(y))(\s_2(x)\s_1(y)+\s_1(x)\s_2(y)) \nonumber \\
&\times&
(\s_1(x)\s_1(y)(\s_2(x)+\s_2(y)+\s_1(x)\s_1(y))-(1+[2]_q)\s_2(x)\s_2(y)).
\nonumber
\eeqa

The polynomial $K_{[m,n]}$ is given by
\beqn
K_{[m,n]}({\bf x},{\bf y})=\prod_{i=1}^{m}\prod_{j=1}^{n}(x_{i}+y_{j})
=\sum_{\la}s_{\la}({\bf x})s_{\hat{\la}}({\bf y}),
\eeqn
which is the kernel of the kinematical residue equations.
Here $s_{\la}$ is the Schur functions and
the summation is taken for the partition $\la=(\la_1,\cdots,\la_m)$,
i.e. non-increasing sequences of non-negative integers under the condition
$\la_1=n$
and $\hat{\la}=(m-\la'_n,\cdots,m-\la'_1)$\cite{M}.
The partition $\la'$ is the conjugate of the partition $\la$.

The polynomials $B_{a[m,n]}(a=1,2)$ are given by
\beqn
B_{1[m,n]}({\bf x},{\bf y})
=\prod_{i<j\leq m}(x_{i}-\omega^2 x_{j})(x_{i}-\omega^{-2} x_{j})
=s_{2\delta_m}({\bf x}),
\eeqn
\beqn
B_{2[m,n]}({\bf x},{\bf y})
=\prod_{i<j\leq n}(y_{i}-\omega^2 y_{j})(y_{i}-\omega^{-2} y_{j})
=s_{2\delta_n}({\bf y}),
\eeqn
which are the kernel of the bound state residue equations.
Here $\delta_m=(m-1,m-2,\cdots, 1)$.

The Schur function can be expressed as \cite{M}
\[
s_{\la}={\rm det}\left(\s_{\la'_i-i+j}\right)_{1\leq i,j\leq l(\la')},
\]
where $l(\la)$ is the length of the partition $\la$.

The polynomials $Q_{[m,n],[1,1]}$ ($[m,n]\neq[1,1]$) have the following forms
\beqn
Q_{[m,n],[1,1]}({\bf x};{\bf y})=(\s_1({\bf x})+\s_1({\bf y}))
(\s_{m-1}({\bf x})\s_n({\bf y})+\s_m({\bf x})\s_{n-1}({\bf y}))
P_{[m,n]}({\bf x};{\bf y}).
\label{PolTh}
\eeqn
 From the stress-energy conservation, it is possible to show that
the polynomials which enter the form factors of the trace of the
stress-energy tensor $\Theta$  are factorized as (\ref{PolTh}).
So the operator ${\cal O}_{[1,1]}$ is identified with $\Theta$.
This fixes the constant $A_{[1,1]}$ to be
\[
A_{[1,1]}=\frac{\pi M^2}{2F^{(min)}_{12}(i\pi)},
\]
where $M$ is the mass of particles.

In this sector, we determined the form factor up to four-body ones.
For example, the explicit form of two-body form factor is
given as
\beqn
F_{12}(\beta)=F_{12}^{\Theta}(\beta)=
\frac{\pi M^2}{2}\frac{F^{(min)}_{12}(\beta)}{F^{(min)}_{12}(i\pi)}.
\eeqn

\subsection{The $[1,0]$-sector and the $[0,1]$-sector}

The solutions in the $[0,1]$-sector are simply obtained from the
$[1,0]$-sector using ${\bf Z}_2$ symmetry.
So we only deal with the $[1,0]$-sector.

First few solutions in the $[1,0]$-sector is given by
\begin{flushleft}
\[
Q_{[1,0]}=A_{[1,0]}.
\]
\[
Q_{[0,2]}=A_{[0,2]}B_{2[0,2]}+HA_{[1,0]}\sigma^{(2)}_{2}.
\]
\[
Q_{[2,1]}=A_{[2,1]}B_{2[2,1]}K_{[2,1]}+HA_{[0,2]}\sigma^{(2)}_{2}K_{[2,1]}+
H^2 A_{[1,0]}\sigma^{(2)}_{1}\sigma^{(2)}_{2}\sigma^{(1)}_{1}.
\]
\beqa
Q_{[4,0]}&=&A_{[4,0]}B_{1[4,0]}+HA_{[2,1]}Q_{[4,0],[2,1]}+
H^2 A_{[0,2]}Q_{[4,0],[0,2]} \nonumber \\
&+&H^3 A_{[1,0]}Q_{[4,0],[1,0]}. \nonumber
\eeqa
\beqa
Q_{[1,3]}&=&A_{[1,3]}B_{2[1,3]}K_{[1,3]}+HA_{[2,1]}Q_{[1,3],[2,1]}+
H^2 A_{[0,2]}Q_{[1,3],[0,2]} \nonumber \\
&+&H^3 A_{[1,0]}Q_{[1,3],[1,0]}. \nonumber
\eeqa
\end{flushleft}
The explicit forms of the polynomials $Q_{[m,n],[m',n']}$ in
the above equations are given by
\beqa
Q_{[4,0],[2,1]}&=&(2+[2]_q)\s_2(-\s_1\s_3^3-\s_1^3\s_3\s_4
+2\s_1\s_2\s_3\s_4+\s_3^2\s_4+\s_1^2\s_4^2) \nonumber \\
&+&\{\s_2^2+(2[2]_q+[2]_q^2)\s_4\}(\s_1^2\s_3^2-
\s_2^2\s_4-2\s_1\s_3\s_4+\s_4^2). \nonumber
\eeqa
\[
Q_{[4,0],[0,2]}=
\s_4\left((2+[2]_q)(\s_4-\s_1\s_3)(\s_4+\s_2^2-\s_1\s_3)
+([3]_q-5)\s_2^2\s_4\right).
\]
\beqa
Q_{[4,0],[1,0]}&=&
\s_4\left((2+[2]_q)\s_2(-\s_3^2-\s_1^2\s_4+\s_2\s_4)\right. \nonumber \\
& & \left. +(\s_2^2+([3]_q+3[2]_q+1)\s_4)(\s_1\s_3-\s_4)\right). \nonumber
\eeqa

\[
Q_{[1,3],[1,0]}=\s^{(3)}_3\s^{(1)}_1\left(\s^{(3)}_1\s^{(3)}_2-
(2+[2]_q)\s^{(3)}_3
\right)\left(\s^{(3)}_2+\s^{(3)}_1\s^{(1)}_1\right).
\]

\[
Q_{[1,3],[0,2]}=\s^{(1)}_1\left(\s^{(3)}_1\s^{(3)}_2-(2+[2]_q)\s^{(3)}_3
\right)
\left((\s^{(3)}_1)^2\s^{(3)}_3+(\s^{(3)}_2)^2\s^{(1)}_1\right).
\]

\[
Q_{[1,3],[2,1]}=\s^{(3)}_1\s^{(3)}_2\left(\s^{(3)}_1\s^{(3)}_2-
(2+[2]_q)\s^{(3)}_3\right)K_{[1,3]}.
\]

The form factor of the fundamental operator is factorized as follows
\beqn
Q_{\mu,\nu}({\bf x},{\bf y})=\s^{(m)}_m({\bf x})\s^{(n)}_n({\bf y})
P_{\mu,\nu}({\bf x},{\bf y}).
\eeqn
The operators ${\cal O}_{[1,0]}$ and ${\cal O}_{[0,1]}$
correspond to these fundamental
operators, and others are composite operators.

Note that $[1,0]$ and $[0,1]$ are fundamental weights of $A_2$ algebra.
The operators labeled by the fundamental weight correspond to the
fundamental operators. From the conservation of the ${\bf Z}_3$ charge,
${\cal O}_{\mu^{(a)}}=\phi_{\bar{a}} \ (a=1,2)$.
Here $\phi_a$ are Affine Toda fields.

The requirement
\[
F_a(\beta)=F^{\phi_{\bar{a}}}_a(\beta)=
<0|\phi_{\bar{a}}(0)|\beta>_a=\frac{1}{\sqrt{2}} \ (a=1,2)
\]
fix the constant $A_{[1,0]}=A_{[0,1]}$ to be $1/\sqrt{2}$.

In $[1,0]$-sector and $[0,1]$-sector, we also determined the
form factors up to four-body ones.
For example, the two-body form factor is given by
\[
F^{\phi_2}_{22}(\beta)=-\sqrt{\frac{3}{2}}\frac{\Gamma}{2\cosh\beta+1}
\frac{F_{22}^{(min)}(\beta)}{F_{22}^{(min)}(\frac{2}{3}\pi i)}.
\]

\resection{The $A_N$ case}

The $A_N$ theory contains $N$ kinds of particles, which are denoted by
$1,\cdots, N$. The anti-particle of $a$ is $\bar{a}=h-a=N+1-a$.
The mass of the particle of type $a$ is given by
$M_a=2M\sin(a\pi/h)$\cite{AFZ}.
As for the case of the $A_2$ theory, the $A_N$ form factors are divided into
$N+1$ sectors. The sector specified by the fundamental sector contains the
fundamental operator,
and the zero sector contains the identity operator and the
stress-energy operator.

Physically, these sectors simply come from
the decomposition of the states
into the different ${\bf Z}_{N+1}$-charge sectors.

For the $A_N$ theory,
the explicit form of $F^{(min)}_{ab}$ can be written as
\[
F^{(min)}_{ab}(\beta)=\prod_{
\scriptstyle x=|a-b|+1 \atop\scriptstyle {\rm step}2}^{a+b-1}
F_x^{(min)}(\beta)
=\prod_{
\scriptstyle x=|a-b|+1 \atop\scriptstyle {\rm step}2}^{h-|a+b-h|-1}
F_x^{(min)}(\beta).
\]
Using $F_x^{(min)}(\beta)F_{2h-x}^{(min)}(\beta)=1$,
we can show $F^{(min)}_{ab}(\beta)=
F^{(min)}_{\bar{a}\bar{b}}(\beta)$.

The monodromy factors $\xi$ and $\lambda$ are given as follows
\[
1/\xi_{ab}(\beta)=\prod_{
\scriptstyle x=|a-b|+1 \atop\scriptstyle {\rm step}2}^{h-|a+b-h|-1}
<x>_{+(\beta)},
\]
\beqn
1/\la^c_{ab;d}(\beta)=
\left\{\begin{array}{lll}
\prod_{\scriptstyle{v(a,b,d)+1} \atop\scriptstyle{{\rm step}2}}
^{b-|a-d|-1}<x>_{+(\beta)} & {\rm for} & a+b+c=h \\
 & & \\
\prod_{\scriptstyle{ v(\bar{a},\bar{b},\bar{d})+1 }\atop
\scriptstyle{{\rm step}2}}^{\bar{b}-|\bar{a}-\bar{d}|-1}<x>_{+(\beta)} &
{\rm for} & a+b+c=2h,
\end{array}\right.
\eeqn
where $v(a,b,d)=|b-d|+b-d-|a+b-d|$.

The explicit form of on-shell three point vertex is
\beqn
(\Gamma^c_{ab})^2=\Gamma^{(h)}<2u^c_{ab}-1>_{+(0)}
\prod_{\scriptstyle x=|a-b|+1 \atop\scriptstyle {\rm step}2}^{u^c_{ab}-3}
\frac{<u^c_{ab}+x>_{+(0)}}{<u^c_{ab}-x>_{+(0)}},
\eeqn
where
\[
\Gamma^{(h)}=\left\{\begin{array}{lll}
2\pi^2/\sin\frac{\pi}{h} &
{\rm for} \ {\rm perturbed \ conformal} \\
 & & \\
-\left(\cos\frac{\pi}{h}-\cos\frac{\pi}{h}(B-1)\right)/\sin\frac{\pi}{h} &
{\rm for} \ {\rm Affine \ Toda.}\end{array}\right.
\]
For the $A_N$ theories, $u^c_{ab}=h-|a+b-h|$ for $a+b+c=0$ mod $h$.
\[
<x>_{+(0)}=\left\{\begin{array}{lll}
-\left(\cos\frac{\pi}{h}x-\cos\frac{\pi}{h}\right)/2\pi^2 &
{\rm for} \ {\rm perturbed \ conformal} \\
 & & \\
\left(\cos\frac{\pi}{h}x-\cos\frac{\pi}{h}\right)/\left(
\cos\frac{\pi}{h}x-\cos\frac{\pi}{h}(B-1)\right) &
{\rm for} \ {\rm Affine \ Toda.}\end{array}\right.
\]

Especially for the Affine $A_2$ theory,
$(\Gamma)^2=\Gamma^{(3)}<3>_{+(0)}$ which agree with the previous result.

The on-shell three point vertex $\Gamma_{ab}^c$ can be expressed as
\beqn
(\Gamma_{ab}^c)^2=
\left\{\begin{array}{lll}
\left(\Gamma_c/\Gamma_a\Gamma_b\right)^2
& {\rm for} & a+b+c=h \\
 & & \\
\left(\Gamma_{\bar{c}}/\Gamma_{\bar{a}}\Gamma_{\bar{b}}\right)^2
 & {\rm for} & a+b+c=2h,
\end{array}\right.
\eeqn
where $\Gamma_a=\prod_{d=1}^{a-1}\Gamma_{1d}^{h-d-1}$.

The $S$-matrices for the $A_N$ theories ($N\geq 3$) contain double poles.
The singularity of form factors are parameterized by the following functions
\beqn
W_{ab}(\beta)=
\prod_{\scriptstyle x=|a-b|+2 \atop\scriptstyle
{\rm step}2}^{u_{ab}^c-2\delta_{a+b,h}}
(x)_+(-x)_+.
\eeqn
In the above factorization, we allow the case $c=0$ i.e. $a+b=h$.
For $a+b=h$, the constant $u_{ab}^c$ is taken that $u_{ab}^0=h$.
Corresponding to the double pole of the $S$-matrices,
additional simple pole appears at (relative) rapidity
$\beta=i\frac{\pi}{h}(u_{ab}^c-2k) \ \
k=1,\cdots,\frac{1}{2}(u_{ab}^c-|a-b|)-1$.
So we must determine these additional pole structure of the form factors.

 From now on, we consider the Affine cases for definiteness.
The one particle form factor $F_a$ is constant.
\[
F_a=<0|\phi_{\bar{a}}(0)|\beta>_a=\frac{1}{\sqrt{2}}.
\]

The elementary two particle form factor is given as
\[
F_{1a}(\beta)=-\Gamma_{1a}^{h-a-1}F_{a+1}
\frac{\sin\theta^{n-a-1}_{1a}}{\cosh\beta-\cos\theta^{h-a-1}_{1a}}
\frac{F^{(min)}_{1a}(\beta)}{F^{(min)}_{1a}(i\theta^{h-a-1}_{1a})} \ \
{\rm for} \ a\neq N.
\]
\[
F_{1N}(\beta)=F^{\Theta}_{1N}(\beta)=
\frac{\pi}{2}M_1^2\frac{F^{(min)}_{1N}(\beta)}
{F^{(min)}_{1N}(i\pi)}.
\]
These two-body form factors play the role of the initial conditions of
the recursion equations.

In order to fix the additional simple pole structure,
we analyze some low order form factors.

Solving the recursion relation for $F_{11a}$, we determined
the explicit form of $F_{2a}$ for $a<N-1$,
\beqa
F_{2a}(\beta)&=&-\Gamma_{2a}^{h-a-2}F_{a+2}
\frac{F^{(min)}_{2a}(\beta)}{F^{(min)}_{2a}(i\frac{\pi}{h}(a+2))} \nonumber \\
& &\times\left\{
\frac{\sin\frac{\pi}{h}(a+2)}{\cosh\beta-\cos\frac{\pi}{h}(a+2)}-
\frac{\sin\frac{\pi}{h}(a+2)}{\cosh\beta-\cos\frac{\pi}{h}a}
\left(
\frac{\cos\frac{\pi}{h}-\cos\frac{\pi}{h}(B-1)}
{\cos\frac{3\pi}{h}-\cos\frac{\pi}{h}(B-1)}\right)\right\}. \nonumber
\eeqa
Also for $a=N-1=\bar{2}$,
\beqa
F_{2\bar{2}}(\beta)&=&\frac{\pi}{2}M_2^2
\frac{F_{2\bar{2}}^{(min)}(\beta)}{F_{2\bar{2}}^{(min)}(i\pi)}
\left\{1-2\sin^2\frac{\pi}{h}\left(
\frac{\cos\frac{\pi}{h}-\cos\frac{\pi}{h}(B-1)}
{\cos\frac{3\pi}{h}-\cos\frac{\pi}{h}(B-1)}\right)\right. \nonumber \\
& & \ \ \ \ \ \ \ \ \ \ \
\times \left.\left(\frac{1}{\cosh\beta-\cos\frac{\pi}{h}(h-2)}+
\frac{1}{1+\cos\frac{\pi}{h}(h-2)}\right)\right\}. \nonumber
\eeqa

One can show that
\beqn
\frac{F_{ab}^{(min)}\left(i\frac{\pi}{h}(u_{ab}^c-2k)\right)}
{F_{ab}^{(min)}\left(i\frac{\pi}{h}u_{ab}^c\right)}=
\frac{F_{l(u_{ab}^c-l)}^{(min)}\left(i\frac{\pi}{h}|a-b|\right)}
{F_{l(u_{ab}^c-l)}^{(min)}\left(i\frac{\pi}{h}u_{ab}^c\right)} \ \ \
k=1,\cdots,\frac{1}{2}(u_{ab}^c-|a-b|)-1,
\label{Fmink}
\eeqn
where $l=k$ for $a+b\leq h$ and $l=\bar{k}$ for $a+b>h$.

Using eq.(\ref{Fmink}), we can see that
\beqn
-i\ {\rm res}_{\beta=i\frac{\pi}{h}a} \
F_{2a}(\beta)=
\Gamma_{11}^{h-2}\Gamma_{1a}^{h-a-1}
F_{1(a+1)}\left(i\frac{\pi}{h}(a-2)\right).
\label{Fta}
\eeqn
For the case of $a=N-1$, the factor $\Gamma_{1(N-1)}^1/\Gamma_{11}^{N-1}$
appears. From the form of $\Gamma_{ab}^c$, it holds that
$(\Gamma_{1(N-1)}^1)^2=(\Gamma_{11}^{N-1})^2$.
In showing eq.(\ref{Fta}), we take the phase
$\Gamma_{1(N-1)}^1/\Gamma_{11}^{N-1}=1$.

In general, it is expected that
\beqn
-i\ {\rm res}_{\beta=i\frac{\pi}{h}(u_{ab}^c-2k)} \
F_{ab}(\beta)
=\Gamma_{lm_1}^{h-l-m_1}
\Gamma_{m_2(m_1-l)}^{h-u_{ab}^c+l}F_{l(u_{ab}^c-l)}
\left(i\frac{\pi}{h}|a-b|\right) \ \ \ k=1,\cdots,m_1-1,
\label{fabk}
\eeqn
where $l=k$ for $a+b\leq h$, $l=\bar{k}$ for $a+b>h$,
$m_1=\frac{1}{2}(u_{ab}^c-|a-b|)$ and $m_2=\frac{1}{2}(u_{ab}^c+|a-b|)$.
And $\Gamma_{1a}^{h-a-1}=\Gamma_{1(h-a-1)}^a$.
We checked the above equation for the case of $a=3$.

The additional simple pole structure is determined as follows:
\beqa
& &-i \ {\rm res}_{\beta'=\beta+i\frac{\pi}{h}(a+b-2k)}
\ F_{abd_1\cdots d_n}
\left(\beta',\beta,\beta_1,\cdots,\beta_n\right) \nonumber \\
&=&\Gamma_{k(a-k)}^{h-a}\Gamma_{(a-k)b}^{h-a-b+k}
F_{k(a+b-k)d_1\cdots d_n}\left(\beta+i\frac{\pi}{h}(b-k),
\beta+i\frac{\pi}{h}(a-k),\beta_1,\cdots,\beta_n\right),
\eeqa
for $a+b\leq h$ and $a\leq b$.

$${\epsfxsize=15.5 truecm \epsfbox{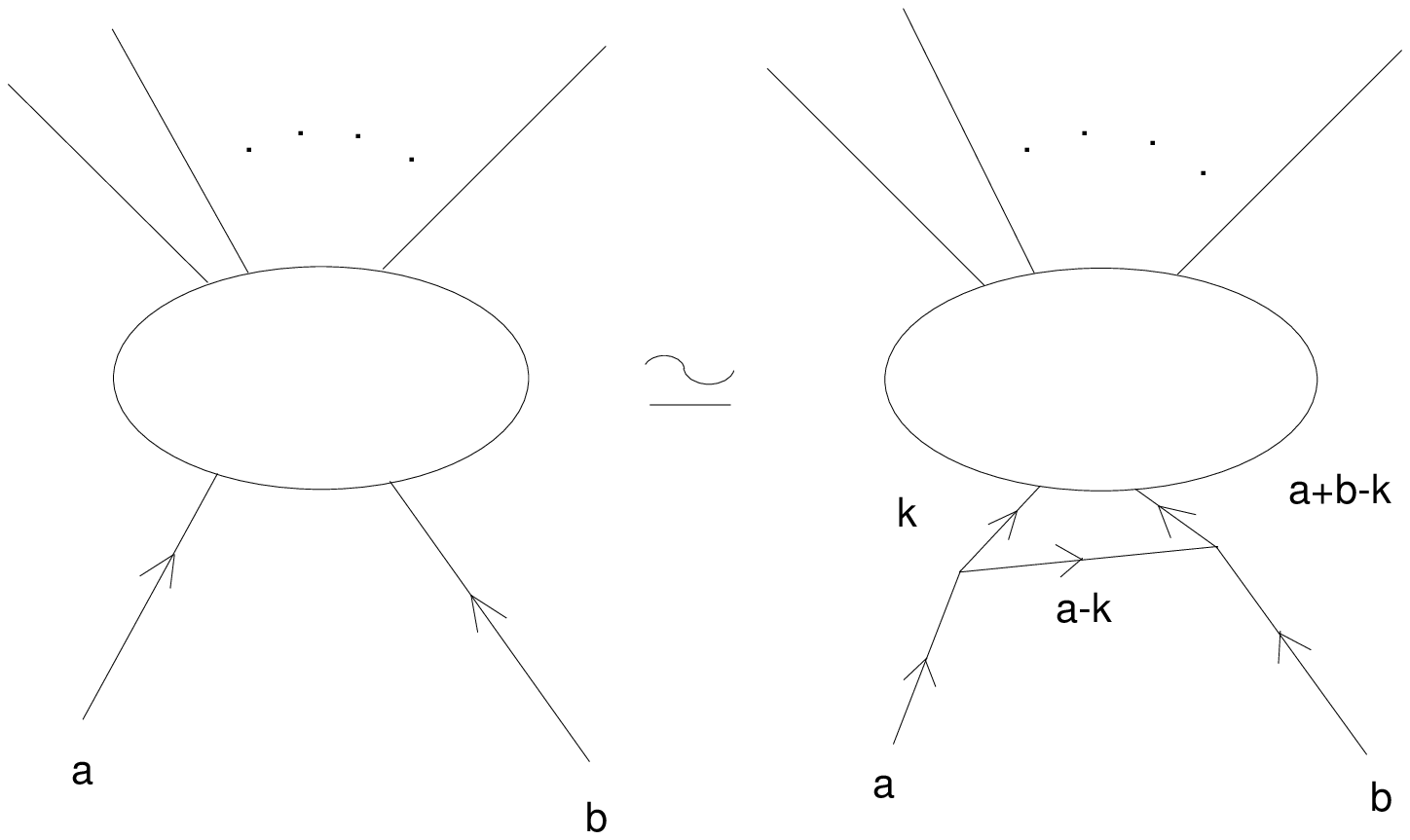}}$$
%\begin{figure}
%\centerline{\psfig{figure=ff2.ps}}
%\end{figure}
\centerline{Figure : The additional pole structure
for $a+b\leq h$ and $a\leq b$.}
\vskip 1.5 truecm

Similar relation holds for other cases.
Above pole structure is similar to that of $A_{2N}^{(2)}$ theories \cite{S}.

Using eq.(\ref{fabk}) recursively, the two-body form factors are determined as
\beqn
F_{ab}(\beta)=\sin\theta_{ab}^c\Gamma_{ab}^cF_{\bar{c}}
\frac{F_{ab}^{(min)}(\beta)}{F_{ab}^{(min)}(i\theta_{ab}^c)}
\sum_{l_0=0}^{\frac{1}{2}(u_{ab}^c-|a-b|)-1}
\frac{B_{ab;l_0}}{\cosh\beta-\cos\frac{\pi}{h}(u_{ab}^c-2l_0)},
\eeqn
where
\beqn
B_{ab;l_0}=\sum_{n\geq 0}(-\frac{1}{2})^n
\sum_{l_1=1}^{l_0-1}\sum_{l_2=1}^{l_1-1}\cdots\sum_{l_n=1}^{l_{n-1}-1}
\frac{\prod_{j=0}^n\sin\frac{\pi}{h}(u_{ab}^c-2l_j)
\left(\Gamma_{l_j(l_{j-1}-l_j)}^{h-l_{j-1}}\right)^2}
{\prod_{j=1}^n\sin\frac{\pi}{h}(l_{j-2}-l_j)\sin\frac{\pi}{h}
(u_{ab}^c-l_{j-2}-l_j)}.
\eeqn
Here $l_{-1}=\frac{1}{2}(u_{ab}^c-|a-b|)$ and $B_{ab;0}=1$.

Also
\beqa
F_{a\bar{a}}(\beta)&=&\frac{\pi}{2}M_a^2
\frac{F_{a\bar{a}}^{(min)}(\beta)}{F_{a\bar{a}}^{(min)}(i\pi)} \nonumber \\
&\times&\left\{1-\sum_{l_0=1}^{m_1-1}B_{a;l_0}
\left(\frac{1}{\cosh\beta-\cos\frac{\pi}{h}(h-2l_0)}+
\frac{1}{1+\cos\frac{\pi}{h}(h-2l_0)}\right)\right\},
\eeqa
where
\beqa
B_{a;l_0}&=&\left(\frac{\sin\frac{\pi}{h}l_0}{\sin\frac{\pi}{h}a}
\Gamma_{l_0(l_{-1}-l_0)}^{h-l_{-1}}\right)^2\sin\frac{2\pi}{h}l_0 \nonumber \\
&\times&\left\{\sum_{n\geq 0}\sum_{l_1=1}^{l_0-1}\sum_{l_2=1}^{l_1-1}
\cdots\sum_{l_n=1}^{l_{n-1}-1}(-\frac{1}{2})^n
\left(\frac{\sin\frac{\pi}{h}a}{\sin\frac{\pi}{h}l_{n-1}}\right)^2
\prod_{j=1}^n\frac{\sin\frac{2\pi}{h}l_j
\left(\Gamma_{l_j(l_{j-1}-l_j)}^{h-l_{j-1}}\right)^2}
{\sin\frac{\pi}{h}(l_{j-2}+l_j)\sin\frac{\pi}{h}(l_{j-2}-l_j)}\right\}.
\nonumber
\eeqa
Here  $l_{-1}={\rm min}(a,\bar{a})$.

\resection{Conclusions and discussion}

We have derived the minimal two-particle form factors for ADE scattering
theories. Using monodromy properties of the building blocks of
minimal form factors, we have determined the functional equations satisfied by
the minimal form factors. The function $\lambda_{ab;d}^c$
will play a key role in
constructing the solutions of the form factor bootstrap equations.

We have determined the parameterization function $W_{ab}$ for the perturbed
conformal theories and for the Affine Toda Field theories.

For the $A_2$ Affine Toda theory, form factors are
derived up to four-body, and
the identifications of the fundamental operators have been done.

For the $A_N$ theories, the form factors are divided into $N+1$ sectors.
To each sector, there corresponds the elementary operator or the stress-energy
operator. This is a generalization of the known result for the sinh-Gordon
theory (i.e. $A_1$ Toda theory) \cite{KM,FMS} to the $A_N$ cases.
For the $A_N$ theories, we have determined the two-particle form factors.
Also, the additional simple pole structure of
form factors has been determined.

The determination of higher order form factors remains to be solved.

Before concluding this article, we state a relation between
roots and equivalence relation of weights.

Dorey's fusion rule for the simply-laced theories \cite{D}
is that the fusion process
$a\times b\rightarrow \bar{c}$ occurs if
\beqn
w^{\xi(a)}\mu^{(a)}+w^{\xi(b)}\mu^{(b)}+w^{\xi(c)}\mu^{(c)}=0,
\label{cabc}
\eeqn
for some integer $\xi(a)$, $\xi(b)$ and $\xi(c)$.
The particle $b$ is the anti-particle of $a$ if
\beqn
w^{\xi(a)}\mu^{(a)}+w^{\xi(b)}\mu^{(b)}=0.
\label{cab}
\eeqn
for some integer $\xi(a)$ and $\xi(b)$.

We define 'fusion base' vector $e_{ab}^{c}=\mu^{(a)}+\mu^{(b)}-\mu^{(c)}$ or
$e_{ab}^{0}=\mu^{(a)}+\mu^{(b)}$ if  eq.(\ref{cabc}) or eq.(\ref{cab})
is satisfied and $e_{ab}^{c}=0$ otherwise.

For the fusion $a\times b\rightarrow\bar{c}$,
if its fusion vector is expanded in simple roots
\beqa
e_{ab}^{c}&=&\mu^{(a)}+\mu^{(b)}-\mu^{(\bar{c})} \nonumber \\
&=&\sum_{d=1}^{r}m_{d}\alpha_d,
\eeqa
then the coefficient $m_d$ take integer values.
This is equivalent to the statement that for $a\times b\rightarrow\bar{c}$,
following condition is necessary:
\[
(C^{-1})^{ad}+(C^{-1})^{bd}-(C^{-1})^{\bar{c}d}\in{\bf Z}.
\]
Here $C$ is the Cartan matrix of the algebra.
For the $E_8$ algebra, all elements of $C^{-1}$ are integers,
so the above condition is always satisfied for any $(a,b,c)$.
Only for the $A_N$ algebra, above condition is also sufficient.

Using the explicit form of the inverse of the Cartan matrix
for the $A_N$ theories,
\[
(C^{-1})^{ab}=\frac{1}{N+1}{\rm min}(a,b)(N+1-{\rm max}(a,b)),
\]
one can show that the above condition is equivalent to $a+b+c=0$ mod $N+1$.

\subsection*{Acknowledgments}

I am grateful to Professor R. Sasaki for instructive suggestions
in the early stage of this work,
and to Professor H. Itoyama for useful discussions.

\end{document}